\documentstyle[eqsecnum,psfig,aps,twocolumn,prl,epsf]{revtex}

\input{psfig}
\begin{document}

\onecolumn

\begin{flushright}
CDF/PUB/EXOTIC/PUBLIC/4460\\
FERMILAB-PUB-98/145-E
\end{flushright}

\begin{center}
{\bf Search for Long-Lived Parents of $Z^0$ Bosons in $p\overline{p}$ 
Collisions at $\sqrt{s}=1.8$ TeV}
\end{center}

\font\eightit=cmti8
\def\r#1{\ignorespaces $^{#1}$}
\hfilneg
\begin{sloppypar}
\noindent
F.~Abe,\r {17} H.~Akimoto,\r {39}
A.~Akopian,\r {31} M.~G.~Albrow,\r 7 A.~Amadon,\r 5 S.~R.~Amendolia,\r {27} 
D.~Amidei,\r {20} J.~Antos,\r {33} S.~Aota,\r {37}
G.~Apollinari,\r {31}
% ...
\end{sloppypar}
\vskip .026in
\begin{center}
(CDF Collaboration)
\end{center}

\vskip .026in
\begin{center}
\r 1  {\eightit Argonne National Laboratory, Argonne, Illinois 60439} \\
\r 2  {\eightit Istituto Nazionale di Fisica Nucleare, University of Bologna,
I-40127 Bologna, Italy} \\
\r 3  {\eightit Brandeis University, Waltham, Massachusetts 02254} \\
\r 4  {\eightit University of California at Los Angeles, Los 
Angeles, California  90024} \\  
\r 5  {\eightit University of Chicago, Chicago, Illinois 60637} \\
\r 6  {\eightit Duke University, Durham, North Carolina  27708} \\
\r 7  {\eightit Fermi National Accelerator Laboratory, Batavia, Illinois 
60510} \\
\r 8  {\eightit University of Florida, Gainesville, FL  32611} \\
\r 9  {\eightit Laboratori Nazionali di Frascati, Istituto Nazionale di Fisica
               Nucleare, I-00044 Frascati, Italy} \\
\r {10} {\eightit University of Geneva, CH-1211 Geneva 4, Switzerland} \\
\r {11} {\eightit Harvard University, Cambridge, Massachusetts 02138} \\
\r {12} {\eightit Hiroshima University, Higashi-Hiroshima 724, Japan} \\
\r {13} {\eightit University of Illinois, Urbana, Illinois 61801} \\
\r {14} {\eightit Institute of Particle Physics, McGill University, Montreal 
H3A 2T8, and University of Toronto,\\ Toronto M5S 1A7, Canada} \\
\r {15} {\eightit The Johns Hopkins University, Baltimore, Maryland 21218} \\
\r {16} {\eightit Institut f\"{u}r Experimentelle Kernphysik, 
Universit\"{a}t Karlsruhe, 76128 Karlsruhe, Germany} \\
\r {17} {\eightit National Laboratory for High Energy Physics (KEK), Tsukuba, 
Ibaraki 305, Japan} \\
\r {18} {\eightit Ernest Orlando Lawrence Berkeley National Laboratory, 
Berkeley, California 94720} \\
\r {19} {\eightit Massachusetts Institute of Technology, Cambridge,
Massachusetts  02139} \\   
\r {20} {\eightit University of Michigan, Ann Arbor, Michigan 48109} \\
\r {21} {\eightit Michigan State University, East Lansing, Michigan  48824} \\
\r {22} {\eightit University of New Mexico, Albuquerque, New Mexico 87131} \\
\r {23} {\eightit The Ohio State University, Columbus, OH 43210} \\
\r {24} {\eightit Osaka City University, Osaka 588, Japan} \\
\r {25} {\eightit Universita di Padova, Istituto Nazionale di Fisica 
          Nucleare, Sezione di Padova, I-35131 Padova, Italy} \\
\r {26} {\eightit University of Pennsylvania, Philadelphia, 
        Pennsylvania 19104} \\   
\r {27} {\eightit Istituto Nazionale di Fisica Nucleare, University and Scuola
               Normale Superiore of Pisa, I-56100 Pisa, Italy} \\
\r {28} {\eightit University of Pittsburgh, Pittsburgh, Pennsylvania 15260} \\
\r {29} {\eightit Purdue University, West Lafayette, Indiana 47907} \\
\r {30} {\eightit University of Rochester, Rochester, New York 14627} \\
\r {31} {\eightit Rockefeller University, New York, New York 10021} \\
\r {32} {\eightit Rutgers University, Piscataway, New Jersey 08855} \\
\r {33} {\eightit Academia Sinica, Taipei, Taiwan 11530, Republic of China} \\
\r {34} {\eightit Texas A\&M University, College Station, Texas 77843} \\
\r {35} {\eightit Texas Tech University, Lubbock, Texas 79409} \\
\r {36} {\eightit Istituto Nazionale di Fisica Nucleare, University of Trieste/
Udine, Italy} \\
\r {37} {\eightit University of Tsukuba, Tsukuba, Ibaraki 315, Japan} \\
\r {38} {\eightit Tufts University, Medford, Massachusetts 02155} \\
\r {39} {\eightit Waseda University, Tokyo 169, Japan} \\
\r {40} {\eightit University of Wisconsin, Madison, Wisconsin 53706} \\
\r {41} {\eightit Yale University, New Haven, Connecticut 06520} \\
\end{center}

\begin{abstract}
\baselineskip 24pt
\begin{small}
We search for new long-lived particles which decay to $Z^0$ bosons
by looking for $Z^0 \rightarrow e^+ e^-$ decays with displaced
vertices.
We find no evidence for parent particles of the $Z^0$
with long lifetimes in 90 pb$^{-1}$ of data from the CDF experiment at Fermilab.
We set a cross section
limit as a function of the lifetime of the parent particle for both a generic
$Z^0$ parent and a
fourth-generation, charge $-\frac{1}{3}$ quark
that decays into $Z^0 b$.\\

\end{small}
\end{abstract}

\draft
%\twocolumn
%\narrowtext
\baselineskip 24pt

In the standard model, there are no particles with mass above 20 GeV and
lifetime greater than $10^{-20}$ seconds.  In particular, there are no metastable
particles that decay into a $Z^0$ boson.  In $p\bar{p}$ collisions, the 
$Z^0$ can be produced either in the primary interaction through quark-antiquark
annihilation or possibly from a neutral-current decay of the short-lived top
quark ($t\rightarrow Z^0c$).  By searching for $Z^0\rightarrow e^+e^-$ with
the $e^+e^-$ vertex displaced from the $p\bar{p}$ interaction point, we are
sensitive to non-standard-model sources of the $Z^0$.

There are a number of extensions to the standard model that can 
accommodate a long-lived parent to a $Z^0$.  One class of models contains 
a fourth-generation, charge $-\frac{1}{3}$ $b ^ \prime$ quark.
A $b ^ \prime$ with mass
less than $m_Z/2$ has been ruled out by experiments at the LEP electron-positron
collider\cite{LEP}.                          
A recent analysis by the D$\O$ collaboration has
excluded the existence of a $b^\prime$ with mass $m_Z /2 < m_{b^\prime} < m_Z + m_b$ 
which decays via the flavor changing neutral current, $b^\prime \rightarrow b +
\gamma$~\cite{d0}.
A more massive $b ^ \prime$ could 
decay into a $Z^0$ and a
bottom quark $(b^{\prime} \rightarrow b + Z^0)$ through a loop induced
flavor-changing neutral current~\cite{bprime}.
This is expected to be a 
dominant decay mode for $  m_Z + m_b  < m_{b^\prime}  < m_t , m_{t^\prime}$,
where $m_{t^\prime}$ is the mass of the $t ^ \prime$ quark (the
partner of the $b^\prime$).
This decay may have a small partial width due to the neutral current decay and
the fourth-generation quark mixing angles\cite{short}.
The competing charged current decay mode,  $b ^ \prime \rightarrow
W^- c$, could also have a very small partial width since it
depends on the mixing of quarks separated by two generations.
  This analysis searches for a long-lived $b^\prime$ in the mass
region above the $Z^0$ through the decay chain $b^\prime \rightarrow Z^0 b$ where $Z^0
\rightarrow e^+ e^-$.

Some models of supersymmetry also allow for long-lived particles which
decay
to $Z^0$.  For example, a low-energy symmetry-breaking model~\cite{SUSY}
in which the gravitino is the
lightest stable particle allows for a long-lived parent of the $Z^0$.
This
model predicts that the lightest neutralino, which could decay into a
$Z^0$ and a gravitino, $\tilde{N^0_1} \rightarrow Z^0 + \tilde{G}$,
may have a long lifetime because of the small coupling constant to
the gravitino.

The data used in this analysis were collected with the Collider Detector at 
Fermilab (CDF) during the 1993-95 Tevatron run, and correspond
to an integrated luminosity of 90 pb$^{-1}$ of $p\bar{p}$ collisions at
$\sqrt{s}=1.8$ TeV.  The CDF detector is described in detail 
elsewhere~\cite{CDFdet}.  We describe here only the detector components 
most relevant to this analysis.  The central tracking chamber (CTC), which
is immersed in a 1.4 T solenoidal magnetic field, measures the momenta and 
trajectories of charged particles in the region $|\eta|<1.1$~\cite{coordinate}.
The four-layer silicon vertex detector (SVX)~\cite{SVX}, located just outside
the beam pipe, provides precise tracking in the plane transverse to the beam
direction, giving a track impact parameter relative to the beam line with a
resolution of $(13+40/P_T)~\mu$m,
where $P_T$ is the transverse momentum of the track in GeV/c.  The transverse
profile of the Tevatron beam is circular with an rms width of $\sim35~\mu$m.
Electromagnetic and hadronic calorimeters surround the solenoid.  This
analysis uses the central detector region ($|\eta|<1$), where there is full
tracking efficiency.

\begin{figure}[h]
\center
\begin{minipage}[t]{3.0in}
\epsfxsize=2.9in
\mbox{\epsffile[45 162 522 666]{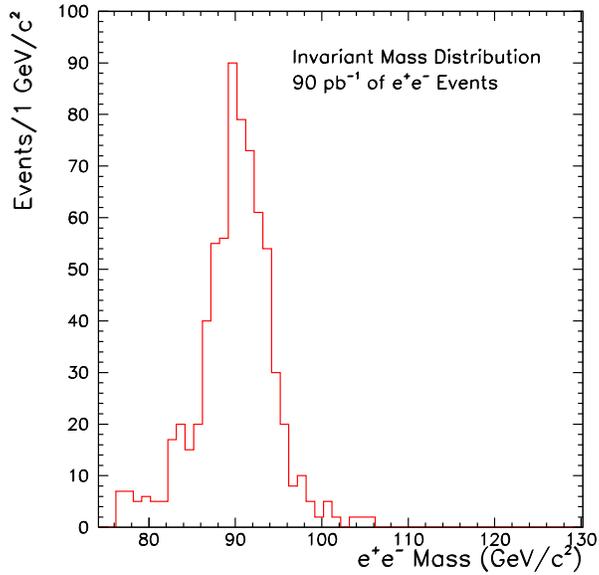}}
\vspace*{0.2in}
\caption{The $e^+e^-$ invariant mass distribution after applying
all cuts.}
\label{fig:mass}
\end{minipage}
\end{figure}

To find a long-lived parent of the $Z^0$, we search for
events containing an electron-positron pair with a mass
consistent with a $Z^0$ and a vertex displaced from the $p\bar{p}$
interaction point.  We begin with a sample of electron-positron pairs,
each lepton having $|\eta|<1$ and $E_T > 20$~GeV~\cite{Sacha}.  The electron
and positron are each required to be isolated, having a total calorimeter
$E_T$ in an $\eta-\phi$ cone of radius 0.4 around the lepton
of no more than 1.15 times the lepton $E_T$.
We also require that the electron-positron invariant mass be in the range
$76.2<M_{ee}<106.2$~GeV$/c^2$ as calculated from the
calorimeter energies and the track directions.    
Because precision
tracking measurements are critical to the determination of the lifetime of the
parent particle, track quality cuts are applied which 
have been optimized using a high-statistics sample of 
$J/\psi \rightarrow \mu^+ \mu^-$ events.  These include minimum numbers of
hits in the SVX and CTC as well as a maximum $\chi^2$ for the track fit.
The electron and positron tracks are fit to a common
vertex, and a good vertex fit is required.
Events are removed if the track pair is within $0.02$ radians of being 
back-to-back ($\Delta\phi$ cut), since nearly collinear tracks have a large uncertainty in the
vertex position.
The invariant mass spectrum of the 703 events that pass all cuts is shown in 
Figure~\ref{fig:mass}.

To search for long-lived $Z^0$ parents, we measure $L_{xy}$, the distance in 
the transverse ($r-\phi$) plane between the $p\bar{p}$ interaction point
and the $e^+e^-$ vertex.  For a long-lived parent, $L_{xy}=\gamma\beta_{xy}ct$,
where $t$ is the proper decay time and $\beta_{xy}$ is the transverse component
of the parent's velocity divided by $c$.  $L_{xy}$ is a signed quantity, the sign
being that of the dot product between two vectors in the transverse plane:
the net $P_T$ of the $e^+ e^-$ pair and the vector pointing from the $p\bar{p}$ interaction
point to the $e^+e^-$ vertex.  $L_{xy}$ significantly less than zero is 
generally due to tracking mismeasurement.  
For standard model direct $Z^0$ production ($q \bar q \rightarrow Z^0
\rightarrow e^+e^-$), we expect $L_{xy} \approx 0$, since
the $Z^0$ lifetime is negligible.
$L_{xy}$ significantly greater than zero suggests that either the $Z^0$ is a 
decay product of a
long-lived parent particle or there is tracking mismeasurement.

\begin{figure}[h]
\center
\begin{minipage}[t]{3.0in}
\epsfxsize=2.9in
\mbox{\epsffile[45 162 522 666]{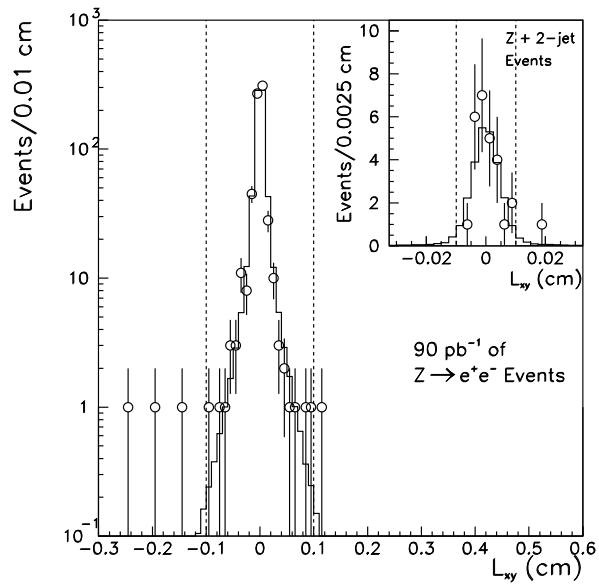}}
\vspace*{0.2in}
\caption {The $L_{xy}$ distribution of the $Z^0$s after applying all cuts.
The data are represented by the circles.  The histogram is the expected $L_{xy}$
distribution for prompt $Z^0$s based on the measured $L_{xy}$ uncertainty in the
event sample. 
The inset shows the distribution after the 2 jet requirement is applied.
The vertical dashed lines separate the prompt and non-prompt regions.}
\label{fig:lxy}
\end{minipage}
\end{figure}

The $L_{xy}$ distribution is shown in
Figure~\ref{fig:lxy} after all of the cuts have been
applied.  The observed distribution is in good agreement with the expected
$L_{xy}$ distribution for prompt $Z^0$s, 
obtained from the $L_{xy}$ uncertainty
measured in each event from propagation of tracking errors. 
Events with large $|L_{xy}|$
that are due to mismeasurement should be symmetric around zero.  
The number of events with $L_{xy}$ significantly less
than zero is thus an effective measure of this background.
To search for long-lived sources, we have examined the events
with $|L_{xy}| > 0.1$ cm, the point beyond which less than one event is
expected from prompt $Z^0$s based on the $L_{xy}$ uncertainty distribution.  
We observe 1
event with $L_{xy} > 0.1$ cm and 3 events with $L_{xy} < -0.1$ cm.
Thus, there is no evidence for a long-lived parent of a
$Z^0$.  We proceed to set limits based on this observation.

The production cross section times branching ratio\cite{sigmabr}
for long-lived parent particle(s) decaying to a $Z^0$
and passing our data selection criteria is calculated by normalizing to the
observed prompt $Z^0$
boson signal.  It can be written as

$$ \sigma \cdot Br \cdot A_X =
    \frac{n_X \cdot A_Z \cdot \sigma_{Z} \cdot Br(Z^0 \rightarrow e^+ e^-)}
 { n_Z \cdot F_{DY} \cdot \epsilon_{\Delta \phi} \cdot \epsilon_{\chi ^ 2}
  \cdot \epsilon_{L_{xy}}
 } $$

\noindent
where  $A_X$ is the
acceptance for finding both the electron and the positron in the geometry of the
detector,
$n_X$ is the number of events seen with a significant
decay length ($>0.1$ cm),
$A_Z = 22 \pm 1 \%$ is
the probability that the $e^-$ and $e^+$ from a directly produced $Z^0$
are in the central part of the
detector,         
$\epsilon_{\Delta \phi}$ is the efficiency of the opening angle cut ($87 \pm 1
\%$ in the direct $Z^0$ sample),
$\epsilon_{\chi ^ 2} = 94 \pm 1 \% $ is the efficiency of the $\chi ^ 2$ cut,
$\epsilon_{L_{xy}}$ is a correction factor for the number of events seen in the
$L_{xy}$ window ($0.1$ cm$<L_{xy}<1.5$ cm),
and  $F_{DY} = 0.96 \pm 0.01$ is a factor to correct
for Drell-Yan contamination in the prompt $Z^0$ sample.
We normalize to the measured $Z^0$ cross section by using
$\sigma_{Z} \cdot Br(Z^0 \rightarrow e^+ e^-)= 231 \pm 12$~pb~\cite{Zcross}
 and $n_Z = 859$, the number of $Z^0$ events left
after the electron pair and tracking
cuts.
For a fixed $\lambda_{xy} \equiv \gamma \beta_{xy} c \tau$, where $\tau$ is 
the lifetime of the parent particle, the efficiency in the $L_{xy}$
window is $\epsilon_{L_{xy}} = e^{\frac{-0.1~{\rm cm}}
{\lambda_{xy}}} - e^{\frac{-1.5~{\rm cm}}{\lambda_{xy}}}$.
                                       
For the $95\%$ confidence level upper limit 
on the cross section, we make the conservative assumption that there is no
background and thus do not perform a background subtraction.
We also conservatively use the opening-angle cut efficiency measured in the
direct $Z^0$ sample.  $Z^0$ bosons from heavy particle decay would generally
be boosted in the transverse direction, thus increasing the cut efficiency.
We use a Poisson distribution based on the one observed event smeared by the
gaussian systematic uncertainties in the acceptance and efficiencies.
We find the $95\%$ confidence level
cross section limit to be
    
$$ \sigma \cdot Br \cdot A_X \leq \frac { 0.36 }
{  
 ( e^{\frac{-0.1~{\rm cm}}{\lambda_{xy}}}
 - e^{\frac{-1.5~{\rm cm}}{\lambda_{xy}}} )
 }~{\rm pb}  $$

\begin{figure}[h]
\center
\begin{minipage}[t]{3.0in}
\epsfxsize=2.9in
\mbox{\epsffile[45 162 522 666]{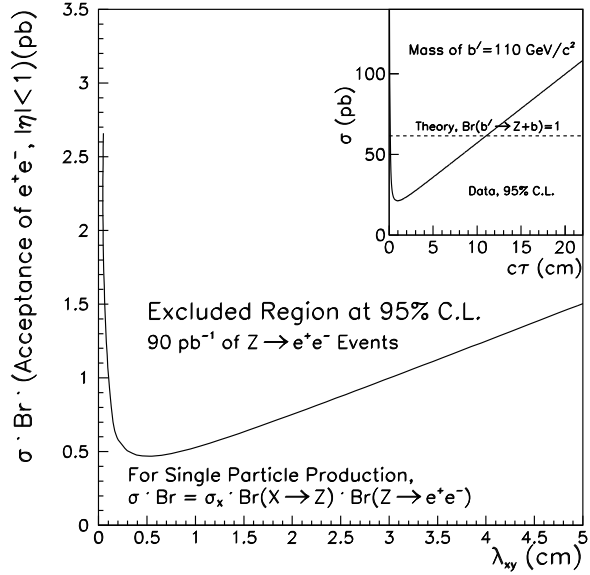}}
\vspace*{0.2in}
\caption {The $95 \%$ confidence level upper cross section limit
for $\sigma \cdot Br $ times the acceptance for an electron-positron pair to be
within the detector
 as a function of fixed $\lambda_{xy}$.
Cross sections above the curve
have been excluded at the $95\%$ confidence level.  The inset shows the
exclusion curve and the theoretical prediction for a $b^\prime$ quark of mass
110 GeV$/c^2$ as a function of its lifetime, assuming 100\% decay into $Z^0b$.}
\label{fig:cross}
\end{minipage}
\end{figure}

\noindent
The cross section limit as
a function of $\lambda_{xy}$ is shown in Figure~\ref{fig:cross}.
                      
A cross section limit can also be determined for $b^ \prime$ pair production.
The $b^ \prime$ quark should have the same production
cross section as a function of mass as the top quark because both are
pair-produced via the strong interaction.  We would also expect to
find several quark jets in the event if a $b ^ \prime$ pair were produced,
for example $q \bar{q}
\rightarrow b ^\prime \bar{b ^ \prime}
\rightarrow b Z^0 \bar b Z ^0
\rightarrow b e ^ + e ^ - \bar b q \bar q$.
We have thus required that there be 2 or more jets
 with $|\eta|<2$ and  $E_T > 10$~GeV.
The $L_{xy}$ distribution for the 27 events surviving
the jet cut is shown in the inset in 
Figure~\ref{fig:lxy}.  The value of $L_{xy}$ above which we expect less than 1
event is now $0.01$ cm.  We find one such event in our data sample.

\begin{figure}[h]
\center
\begin{minipage}[t]{3.0in}
\epsfxsize=2.9in
\mbox{\epsffile[45 162 522 666]{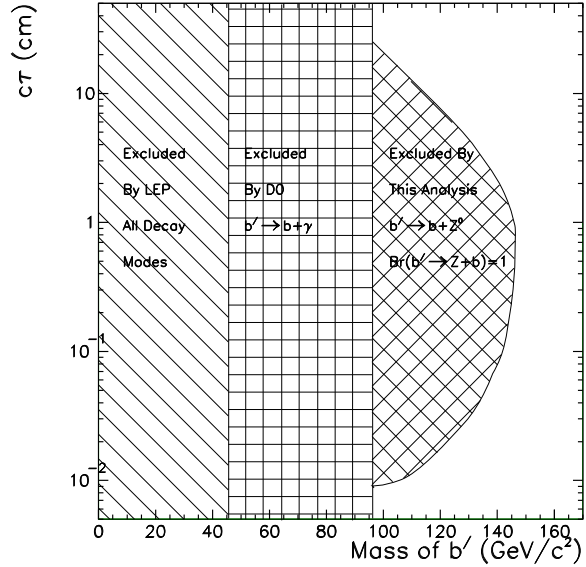}}
\vspace*{0.2in}
\caption {The hatched areas in this plot represent the $95 \%$ confidence-level
regions of $b ^ \prime$ mass and lifetime
that have been excluded.  For $c \tau = 1$ cm, we have excluded up to a mass of
148 GeV$/c^2$.}
\label{fig:exclusion}
\end{minipage}
\end{figure}

The cross section limit for $b ^ \prime$ pair production is given by

$$ \sigma_{b ^ \prime \bar{b^\prime}} \cdot Br(b ^ \prime\bar{b^\prime}
\rightarrow Z^0+X \rightarrow e^+ e^-+X) =$$
 
$$    \frac{n_{b ^ \prime} \cdot A_Z \cdot \sigma_{Z} \cdot
 Br(Z^0 \rightarrow e^+ e^-)}
 { n_Z \cdot F_{DY} \cdot A_{b ^ \prime}  \cdot \epsilon_{jet}
 \cdot \epsilon_{\Delta \phi} \cdot
\epsilon_{\chi ^ 2} \cdot \epsilon_{L_{xy}} \cdot F_I
 } $$

\noindent
where $n_{b^{\prime}}$ is the number of events seen with $L_{xy}>0.01$~cm.
If $b^{\prime}$ always decays into $Z^0b$, the probability that at least one
$Z^0$ decays into $e^+e^-$ is 0.0662.  The quantities 
$\epsilon_{\Delta \phi}$ (the efficiency of the opening angle criterion),
 $\epsilon_{jet}$ (the efficiency of the jet requirement), and
$A_{b ^\prime}$ (the probability of observing an electron and a 
positron in the detector fiducial volume) all depend on the
mass of the $b ^ \prime$.  We use the Herwig Monte Carlo to estimate these
quantities as a function of the $b^\prime$ mass  \cite{herwig}.
We use $\gamma \beta_{xy}$ distributions for
the $b^\prime$ from the Monte Carlo to estimate
$\epsilon_{L_{xy}}$.
This efficiency depends on the mass of the $b ^ \prime$ and the
lifetime, which is a function of the fourth-generation mixing angles between 
the quarks. 
For a particular lifetime, we find $\epsilon_{L_{xy}}$ by calculating
$ e^{\frac{-0.01~{\rm cm}}{\lambda_{xy}}}
 - e^{\frac{-1.5~{\rm cm}}{\lambda_{xy}}} $
for each event and averaging the entire Monte Carlo sample.  
We also include in the calculation of the $b^\prime$ cross
section a factor $F_I = 0.92 \pm 0.05$
that corrects for the reduced electron isolation efficiency due to the expected 
jets in a $b ^ \prime$ event.
The excluded lifetimes for a $b^\prime$ of mass $110$ GeV$/c^2$ are shown in the
inset in Figure~\ref{fig:cross}.
The excluded region of the $b^\prime$ mass versus lifetime plane is shown 
in Figure~\ref{fig:exclusion}
using the theoretical cross sections in \cite{topcross} and the assumption
that $Br(b^{\prime}\rightarrow Z^0+b) = 100\%$.  
             
In conclusion, we find no evidence for new particles with a
long lifetime decaying to $Z^0$ bosons.  We set $95 \%$ confidence-level
cross section upper limits on new particle production as a function of
 $\lambda_{xy}$.
A range in mass and lifetime for a fourth generation $b^\prime$ quark decaying
to $Z^0b$ has been excluded.

We thank the Fermilab staff and the technical staff of the participating
institutions for their contributions.  This work was supported by the U.S.
Department of Energy and National Science Foundation; the Italian Istituto
Nazionale di Fisica Nucleare; the Ministry of Science, Culture, and Education
of Japan; the Natural Sciences and Engineering Research Council of Canada; the
National Science Council of the Republic of China; and the A. P. Sloan
Foundation.

\clearpage

\begin{thebibliography}{99}
\bibitem{short}
A search for a short-lived $b^{\prime}$ quark by looking at $Z^0$ plus three
or more jets, one of which is identified as a $b$ quark, is in progress within
CDF.
\bibitem{coordinate}
The coordinate system at CDF has been defined so the z-axis is the proton
beam direction, $\theta$ is the angle from the positive z-axis, $\phi$ is the
azimuthal angle, and r is the
distance from the z-axis.  Pseudorapidity is defined by $\eta \equiv
-\ln \tan{\frac{\theta}{2}}$.  Transverse momentum is defined by $P_T \equiv P
\sin{\theta}$, and transverse energy deposited in the calorimeters is $E_T
\equiv E \sin{\theta}$.
\bibitem{sigmabr}
$\sigma \cdot Br$
varies for the particular decay process.  For single particle production,
$\sigma \cdot Br = \sigma_X \cdot Br(X \rightarrow Z^0 + Anything) \cdot
Br(Z^0 \rightarrow e^+ e^-)$.  For pair production, the branching ratio
will depend on the probability of producing at least one $Z^0$ that decays
to $e^+ e^-$.
\end{thebibliography}
\end{document}